\documentclass[aps,
pre,
superscriptaddress,
amsmath,amssymb,
reprint,
showkeys,
]{revtex4-2}

\usepackage{algorithm}
\usepackage{algpseudocode}
\usepackage{array}
\usepackage{mathtools}
\usepackage[dvipsnames]{xcolor}
\usepackage{lipsum}
\usepackage{tikz}
\usepackage{wrapfig}
\usepackage[T1]{fontenc}
\usepackage[utf8]{inputenc}
\usepackage{amsmath}
\usepackage{amssymb}
\usepackage{graphicx}
\usepackage{multirow}
\usepackage{blkarray,booktabs,bigstrut}
\usepackage{placeins}
\usepackage[
citecolor=blue,
colorlinks,
linkcolor=blue,
urlcolor=blue,
]{hyperref}

\usepackage{xcolor}
\usepackage{dcolumn}
\usepackage{bm}
\usepackage{tabularx,booktabs}
\usepackage{multirow}
\usepackage{float}
\usepackage[normalem]{ulem}

\begin{document}
	
	
	\title{{Inferring to C or not to C: Evolutionary games with Bayesian inferential strategies}}
	%
	\author{Arunava Patra}
	\email{arunava20@iitk.ac.in}
	\address{
		Department of Physics,
		Indian Institute of Technology Kanpur,
		Uttar Pradesh 208016, India
	}
	
	\author{ Supratim Sengupta }
	\email{supratim.sen@iiserkol.ac.in (Corresponding author)}
	\address{
		Department of Physical Sciences,
			Indian Institute of Science Education and Research Kolkata, Mohanpur Campus, West Bengal 741246, India
	}
	\author{Ayan Paul }
	\email{a.paul@northeastern.edu (Co-corresponding author)}
	\address{Department of Electrical and Computer Engineering, Northeastern University, Boston MA 0211,USA
	}
	\author{Sagar Chakraborty}
	\email{sagarc@iitk.ac.in}
	\address{
		Department of Physics,
		Indian Institute of Technology Kanpur,
		Uttar Pradesh 208016, India
	}
	\begin{abstract}
		Strategies for sustaining cooperation and preventing exploitation by selfish agents in repeated games have mostly been restricted to Markovian strategies where the response of an agent depends on the actions in the previous round. Such strategies are characterized by lack of learning. However, learning from accumulated evidence over time and using the evidence to dynamically update our response is a key feature of living organisms. Bayesian inference provides a framework for such evidence-based learning mechanisms. It is therefore imperative to understand how strategies based on Bayesian learning fare in repeated games with Markovian strategies. Here, we consider a scenario where the Bayesian player uses the accumulated evidence of the opponent’s actions over several rounds to continuously update her belief about the reactive opponent's strategy. The Bayesian player can then act on her inferred belief in different ways. By studying repeated Prisoner's dilemma games with such Bayesian inferential strategies, both in infinite and finite populations, we identify the conditions under which such strategies can be evolutionarily stable. We find that a Bayesian strategy that is less altruistic than the inferred belief about the opponent's strategy can outperform a larger set of reactive strategies, whereas one that is more generous than the inferred belief is more successful when the benefit-to-cost ratio of mutual cooperation is high. Our analysis reveals how learning the opponent's strategy through Bayesian inference, as opposed to utility maximization, can be beneficial in the long run, in preventing exploitation and eventual invasion by reactive strategies.
	\end{abstract}
	\maketitle
	\section{Introduction}
Every organism, from a microbe to a human, reacts to its environment to various degrees of complexity depending on how much cognitively capable it is. Some learn to react to reap more benefit compared to  the others. Such optimal learning is selected over the generations, and finally the progenies are not observed to be spending almost any time in learning to optimally react to the environment; they react instinctively: What was a learnt strategy for the ancestors, is now a genetically-hardwired instinct for the selected descendants~\cite{Baldwin1896, Harley1981, Ancel1999, hinton1987learning, Badyaev2009,Fontanari2017, Morgan2020}. 
	
Specific reaction to a situation is contingent on the information about the situation that the organism gathers. In more formal terms, the organism develops a belief about the state of the environment and its strategy on how to react is based on this belief which can be expressed as a probability distribution over the possible states of the environment. Note that, even before observing the situation, the organism has some belief (called prior belief) that transforms to an updated belief (called posterior belief) in the light of new information. How a posterior belief emerges from a prior belief is dependent on the exact nature of the update rule. 

The optimal update rule itself needs to be learnt and, if it is evolutionarily beneficial, then it would be passed on to the progenies. It has been argued that Bayesian updating ~\cite{Bayes1763, Jaynes2003} is a requirement of evolutionary optimality. While Bayesian updating is not the only form of updating seen in the natural world, there are plenty of examples \cite{McNamara2006, PrezEscudero2011} where the belief updating occurs in accordance with Bayes rule. In passing, we point out that since a player's actions should be probabilistically coherent so as not to incur any loss (say, via the Dutch book), the Bayesian updating is supposedly the updating rule that is a rational player’s normative choice. Whether Bayesian updating is indeed a requirement of rationality, remains a much debated topic ~\cite{Okasha2013}.
	
Of course, the simplest strategy for reacting is to randomly adopt an action (out of a set of all possible implementable actions) even without observing the situation at hand. A more sophisticated strategy would involve choosing an action with some probability, based on knowledge about the particular state of the environment in the immediate past. This may be called a reactive strategy~\cite{Axelrod1981, Nowak1990}. In the context of evolutionary games~\cite{Smith1973, Smith1982}, the reactive strategies are most often encountered in the study of evolution of cooperation~\cite{Axelrod1981, Milinski1987} which seeks to address the question of how altruistic behaviour can be sustained~\cite{Trivers1971} despite involving a cost which makes the altruistic phenotype less suitable for selection over its selfish counterpart. The environment of a player playing an evolutionary game is the entire population consisting of all other players she can interact with.

A lot of work carried out over the last three decades have shed light on this problem by identifying the critical factors that can facilitate not just the survival but also the dominance of cooperators in an evolving population. A vast amount of literature on evolutionary game theory have focused on either reactive and memory-one strategies~\cite{Nowak1993, Nowak2006, Stewart2013,Stewart2014, Stewart2016, Park2022}, population structure~\cite{Nowak1992, Brauchli1999, Santos2005, Ohtsuki2006, Szab2007}, player's reputation together with social norms~\cite{Milinski2002, Santos2016} and strategy update based on the pairwise comparison rule~\cite{Szab1998, Hauert2005, Traulsen2007} to understand the evolution of cooperation in diverse scenarios. Most of the work in this domain can be broadly classified under the categories of direct and indirect reciprocity~\cite{Boyd1989, Nowak1998, Nowak2005,Ohtsuki2004,Imhof2009,Baek2016,Hilbe2017,Hilbe2018,Schmid2021} in either well-mixed or structured populations. Such models focus on identifying strategies that depend only on the strategy of the interacting partner or the strategy of both the focal player and her interacting partner (in the case of memory-one strategies) in the last round. Even when strategy update is carried out using the pairwise comparison model, preferential selection of strategies is based on a single factor, namely payoff of a randomly selected neighbour. Such models constitute a very narrow set of possible strategies that players may employ in deciding whether to be altruistic or selfish. More importantly, these models are characterized by lack of learning based on new evidence that results from repeated interactions, since the strategy-update rule is fixed at the outset. 
	
In view of the above, it is quite natural that one takes up the endeavour of not only considering different strategy update rules~\cite{McAvoy2019, Minjae2021, PVRS2020, Pal-Sengupta2022, McAvoy-Hilbe2022}, that are not just dependent on the payoff ~\cite{PVRS2020}, but also allow for the possibility that individuals may learn the appropriate rule (model for strategy update) from the experience gained through repeated interactions. In other words, a more complex strategy could involve adopting a strategy contingent on the belief about environmental state of player. This belief could further be continually updated based on the actual sequence of states observed over the past times till the true frequencies of environmental states is learnt. Such a strategy may be called a learning strategy.

If the learning strategy specifically calls for Bayes' updating rule, it may be called a Bayesian strategy. It is natural to ponder whether a Bayesian strategy is evolutionary optimal compared to a reactive strategy. To the best of our knowledge, this important question has not yet been investigated in the context of evolutionary games (see, however,~\cite{Minjae2021}). Unlike pairwise-comparison or reinforcement learning mechanisms~\cite{Arthur1993, Brgers1997, Rustichini1999, Laslier2001, Hopkins2002,Beggs2005, Golman2009,Ianni2014} like Bush-Mosteller \cite{ Bush-Mosteller1951, Karandikar1998, Posch1999, Macy-Flache2002,Masuda-NakamuraJTB2011, Tanabe-MasudaJTB2011, Pal-Sengupta2022}, the Bayesian strategy does not attempt to optimise the received payoff in every round but attempts to {\it infer} the reactive strategy of her opponent based on accumulating knowledge of the opponent's actions.  In the process of doing so, the accumulated payoff of the Bayesian player in the long term can exceed the payoff of the opponent using a reactive strategy, under certain circumstances.

Within the paradigm of repeated games, we consider a player employing a Bayesian strategy against an opponent, employing an unknown but fixed reactive strategy. The Bayesian player then tries to infer the reactive strategy (i.e., the probabilities $p^{(r)}$ and $q^{(r)}$ of cooperation of the reactive player when her opponent cooperated (played $C$) and defected (played $D$) respectively in the last round). The Bayesian player uses the action of her reactive interacting partner in each round to update her own beliefs about the ($p^{(r)},q^{(r)}$) values of the partner's reactive strategy using Bayes' rule. She then utilizes her updated belief ($p^{\rm max},q^{\rm max}$) about the interacting player's ($p^{(r)},q^{(r)}$) values to tune her own response in three different ways (see Methods section for details). In the first scenario, we assume that the Bayesian player adopts her updated belief ($p^{\rm max},q^{\rm max}$) about the ($p^{(r)},q^{(r)}$) values of the reactive opponent as her own in the next round. We call such a strategy Bayesian tit-for-tat (BTFT). This ability, called probability matching, of an organism to choose a behaviour with a probability equal to the Bayesian-estimated maximum a posteriori (MAP) probability has been well-documented in several animal behaviour experiments~\cite{Luttbeg1996, Mazalov1996, Welton2003, JValone2006,Biernaskie2009,McNamara2006, PrezEscudero2011}. In an alternative scenario, the Bayesian player adopts a strategy that is less reciprocal than BTFT, i.e., she chooses to cooperate with a probability that is {\it less} than the inferred probabilities of cooperation ($p^{\rm max},q^{\rm max}$)  of the BTFT player. In another alternative scenario, the Bayesian player adopts a strategy that is more generous than BTFT, i.e., she chooses to cooperate with a probability that is {\it more} than the inferred probabilities of cooperation ($p^{\rm max},q^{\rm max}$)  of the BTFT player. The comparative effects of reciprocity and generosity in the evolution of cooperation is a problem of interest for the researchers of evolutionary game theory~\cite{Kurokawa2010,2017_park_pre,Kurokawa2018,Sadhukhan2022}.

Under the Darwinian paradigm all the players adopt an action (or a probabilistic mixture of actions) that makes the population evolutionarily resilient against invasion by a mutant with a different action. Such a robust action is known as a evolutionarily stable strategy (ESS)~\cite{Smith1973, Taylor1978}. Here, we study a repeated two-player game between a Bayesian player and a reactive player and ascertain the conditions for strategies to be ESS. Our results show that the success of Bayesian strategies in preventing invasion by opponents with reactive strategies depend on the manner in which the Bayesian player responds to her continuously updated belief about the opponent's reactive strategy.  
	
		\begin{figure*}
		\centering
		\includegraphics[scale=0.9]{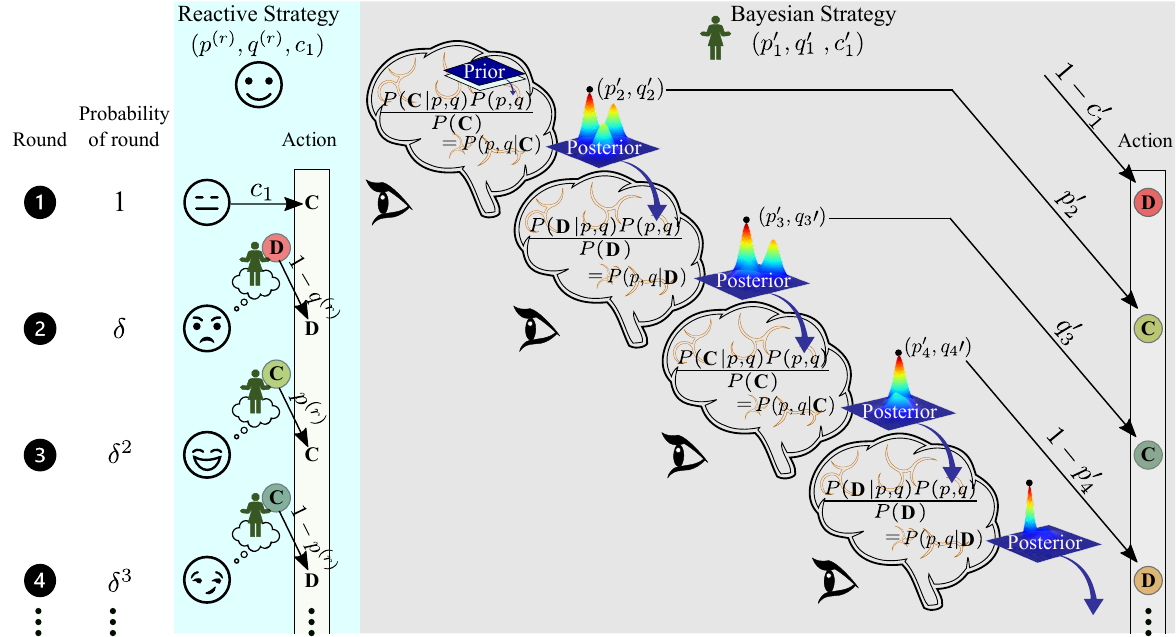}
		\caption{Schematic diagram showing how a Bayesian player updates her belief about an opponent's reactive strategy ($p^{(r)}$, $q^{(r)}$) that is unknown to her. The Bayesian player's objective is to infer the strategy of the reactive player from the latter's actions over several rounds. The Bayesian player is depicted in green and the reactive opponent is denoted by emoticon. In the first round, the reactive player's action is $C$ while the Bayesian player's action is $D$. In response, the reactive player defects ($D$) in round 2 (with probability ($1-q^{(r)}$)). The Bayesian player uses the reactive opponent's action at the end of round 1 as evidence to update her belief ($p^{\prime}_{2},q^{\prime}_{2}$) about the reactive opponent's strategy using Bayes rule. The Bayesian player then uses her updated belief ($p^{\prime}_{2},q^{\prime}_{2}$) to cooperate ($C$) with the reactive opponent (with probability $p^{\prime}_{2}$ in round 2). After round 2, where the reactive player defects, the Bayesian player again uses Bayes rule to update her belief by determining the maximum ($p^{\prime}_{3},q^{\prime}_{3}$) of a new posterior distribution that uses the posterior distribution estimated at the end of round 1 as prior. She then uses this inferred belief to cooperate ($C$) with the reactive opponent with probability $q^{\prime}_{3}$ in response to the reactive opponent's action ($D$) in round 2. This updating process continues over rounds (see Sec.~\ref{sec:II} for details) with the prior being updated to the posterior distribution at the beginning of each round.} 
		\label{fig: Schematic}
	\end{figure*}
\section{Model}
\label{sec:II}
We consider a repeated two-player, two-action game where each player can play with a reactive strategy or a Bayesian strategy. The two actions of the underlying game are taken as cooperation ($C$) and defection ($D$). The reactive strategy is defined by ($p^{(r)}$, $q^{(r)}$, $c_1$). Here $p^{(r)}$ and $q^{(r)}$ are the probabilities of cooperation in the current round of the game given her opponent respectively cooperated and defected in the previous round, while $c_1$ is the reactive player's {\it initial} (first round) probability of cooperation. 
	
A player with a Bayesian strategy attempts to infer the true probabilities of cooperation ($p^{(r)}$ and $q^{(r)}$) of the opponent by taking into account the opponent's actions over several rounds of the game. In order to do so, the Bayesian player has to continuously update her belief about the reactive strategy on the basis of evidence obtained in the form of action ($A\in\{C,D\}$) of the opponent in each round. This update is done using Bayes' rule which requires a prior probability distribution of the beliefs of  the Bayesian inferential player about the reactive strategy adopted by her opponent. We assume a {\it uniform} prior distribution $P_{1}(p,q)$  and at the end of each round $n$ the Bayesian player updates her {\it belief} $(p,q)\in[0,1]\times[0,1]$ about her opponent's strategy from the estimated posterior probability distribution, 
		\begin{equation}
		P_{n}(p,q|A_n)=\frac{P(A_{n}|p,q)P_{n}(p,q)}{\sum_{0}^1 \sum_{0}^1 P(A_{n}|p,q)P_{n}(p,q)}.\label{eq:Bayes}
		\end{equation}
The subscript $n$ alludes to the quantities estimated in the $n$th round. 

The Likelihood, $P(A_{n}|p,q)$, is chosen to be
	\begin{equation}
	P(A_n|p,q) =
	\begin{cases}
	pc^{\prime}_{n-1} +q(1-c^{\prime}_{n-1})~ \text{if}~A_n=C, & \\
	(1-p)c^{\prime}_{n-1} +(1-q)(1-c^{\prime}_{n-1})~\text{if}~A_n=D, & 
	\end{cases}  
	\label{eq:likelihood_gen}   
	\end{equation}
where $c'_n$ is the probability of cooperation of the Bayesian player in $n^{th}$ round. Probabilities $p$ and $q$ are sampled from the set of all possible values lying between 0 and 1. That this choice of likelihood function is apposite is clear from the fact~\cite{Nowak1990,Hofbauer1998} that a reactive player with $p^{(r)}=p$ and $q^{(r)}=q$ plays action $A$ in $n$th round with probability given by Eq.~(\ref{eq:likelihood_gen}) when the opponent's cooperation probability is $c'_{n-1}$ in the previous round. For round $n=1$, the expression for likelihood requires specification of $c'_0$ for which we set $c^{\prime}_{0}=c^{\prime}_{1}$, the initial probability of cooperation of the Bayesian player. 
	
The Bayesian player determines the global maximum $(p^{ \rm max}_{n},q^{\rm max}_{n})$, i.e., the {\it maximum a posteriori} (MAP) of the updated posterior distribution $P_n(p,q|A_n)$ at the end of $n^{th}$ round, which then constitutes her updated belief about the true ($p^{(r)}$,$q^{(r)}$) values of her reactive opponent. If the posterior distribution has multiple maxima, the Bayesian player selects any of them at random. Subsequently, she uses a function of her updated belief as her new strategy in the next round that is
\begin{equation}
	(p'_{n+1},q'_{n+1})=\left(f(p^{\rm max}_{n},q^{\rm max}_{n}),g(p^{\rm max}_{n},q^{\rm max}_{n})\right).
\end{equation}
In this paper, we consider three possible functional forms of $(f,g)$ as defined later. 

Thus, the evolving strategy of the Bayesian player is determined by her {\it belief} ($p$,$q$) about the reactive opponent's strategy. This update is recursive; the posterior distribution calculated in the $n$th round becomes the prior in the $(n+1)$th round, $P_{n+1}(p, q) = P_{n}(p, q|A_n)$, $A_n$ being the action of reactive player in the $n^{th}$ round. Through this updating process, the Bayesian player eventually infers the reactive opponent's true strategy. Had the opponent also been a Bayesian player, even then the focal Bayesian player would implement the Bayes' rule as given above to update her strategy; even though there exists no true belief in any such interaction.

	\subsection{Strategies}
The Bayesian player can adopt her strategy in many ways using her belief, but below we describe three intuitive strategies, viz., Bayesian Tit-for-tat, non-reciprocal Bayesian Tit-for-tat, and generous Bayesian Tit-for-tat.  
	
\textit{Bayesian Tit-for-tat (BTFT):} An obvious choice of a new strategy would be one where the Bayesian player just adopts her updated belief as her new probabilities of cooperation in the next round. Mathematically,
		\begin{subequations}
			\begin{eqnarray}
			\label{eqn:non_recp}
			p'_{n+1}=f(p^{\rm max}_{n},q^{\rm max}_{n})&\equiv& p^{\rm max}_{n}, \\
			\label{eqn:non_recq}
			q'_{n+1}=g(p^{\rm max}_{n},q^{\rm max}_{n})&\equiv& q^{\rm max}_{n}.
			\end{eqnarray}
		\end{subequations}
		We therefore call such a strategy the Bayesian TFT (BTFT) strategy.
		
\textit{Non-reciprocal Bayesian Tit-for-tat (NBTFT):} In some cases, the player may be less cooperative than suggested by her own Bayesian updated belief ($p$,$q$) about her opponent's strategy, due to her own self-interest. We call such a class of strategies non-reciprocal BTFT (NBTFT) strategies. If the parameter $\nu$ measures the extent to which the Bayesian player is less cooperative than a BTFT player, then her updated strategy in the subsequent round becomes
		\begin{subequations}
			\begin{eqnarray}
			\label{eqn:non_recp}
			p'_{n+1}=f(p^{\rm max}_{n},q^{\rm max}_{n})&\equiv& p^{\rm max}_{n}(1-\nu), \\
			\label{eqn:non_recq}
			q'_{n+1}=g(p^{\rm max}_{n},q^{\rm max}_{n})&\equiv& q^{\rm max}_{n}(1-\nu),
			\end{eqnarray}
		\end{subequations}
		where $\nu\in[0,1]$. If $\nu=0$, then the player is reciprocal, and we get back the BTFT strategy. If $\nu=1$, the player is maximally non-reciprocal, which amounts to the Always-defect (ALLD) strategy.
		
\textit{Generous Bayesian Tit-for-tat (GBTFT):} On the other hand, the player may want to decide to defect less than suggested by her belief about the opponent's strategy. We call such a class of strategies Generous BTFT (GBTFT). If the generosity parameter $\gamma$ measures the extent to which the Bayesian player is less selfish than a BTFT player, then her updated probabilities of defection in the subsequent round are
		\begin{subequations} 
			\begin{eqnarray}
			\label{eqn:non_genp}
			1-p'_{n+1}=(1-\gamma)(1-p^{max}_{n}), \\
			\label{eqn:non_genq}
			1-q'_{n+1}=(1-\gamma)(1-q^{max}_{n}),
			\end{eqnarray}
		\end{subequations}
		where $\gamma \in [0,1]$. 
  When $\gamma=0$, the BTFT strategy is recovered. On the other hand, if $\gamma=1$, the player is maximally generous, which amounts to the Always-cooperate (ALLC) strategy.
	
	\subsection{Payoff}
	We consider the underlying game to be a prisoner's dilemma (PD) game with two parameters~\cite{HAMILTON_1964,1971_Trivers}, $b$ and $c$. If a player cooperates, she incurs a cost $c$ and gets the benefit of either $b$ (if the opponent cooperates) or $0$ (if the opponent defects). Whereas, if a player defects, she gets a benefit $b$ without incurring any cost when the opponent cooperates, but gets nothing when the opponent defects. For convenience, we re-parameterise the resulting payoff matrix by dividing all payoff elements by $c$. The effective payoff matrix, thus, is
    \begin{equation}	
	{\sf U}\equiv
	\begin{blockarray}{ccc}
	&  C & D &  \\
	\begin{block}{c[cc]}
	C~ & r-1 & -1\\
	D~ & r & 0\\ 
	\end{block}
	\end{blockarray}~,
	\end{equation}
 where $r\equiv{b}/{c}$ is benefit-to-cost ratio for cooperating.
	The payoff elements satisfy the repeated PD game conditions: $r>r-1>0>-1$ and $2(r-1)>r-1$.
	
The shadow of the future~\cite{2003_Skyrms} is practically inevitable in repeated games. The occurrence of any subsequent round need not be a certainty but it may have a probability $\delta$ between zero and one associated with it. The probability $\delta$ is also called discount factor because one can equivalently consider the payoff in every subsequent interaction to be discounted by a multiplicative factor $\delta$. Clearly, the probability that the game goes on for $n$-rounds is $\delta^{n-1}$. Hence, the accumulated payoff of the player at the end of game, i.e. when $n=n_f$, is 
	\begin{equation}
	\pi(\delta,n_f)=\sum_{n=1}^{n=n_f} u_n\delta^{n-1}.
	\label{eq:Payoff_calculation}
	\end{equation}
Here $u_n$ $\in$ $\{r,-1,r-1,0\}$ is the payoff of the focal player in $n^{th}$ round. 

As players in a population interact at random, a focal player can meet with either a reactive player or a Bayesian player. Therefore, three distinct types of interaction are present in the population: reactive-reactive, reactive-Bayesian, and Bayesian-Bayesian. Hence one can write the payoff matrix,
 \begin{equation}	
	{\sf\Pi}\equiv~
	\begin{blockarray}{ccc}
	&  \textrm{Reactive}\,(R) & \textrm{Bayesian}\,(B) &  \\
	\begin{block}{c[cc]}
	\textrm{Reactive}\,(R)~ & \pi_{RR} & \pi_{RB}\\
	\textrm{Bayesian}\,(B)~ & \pi_{BR} & \pi_{BB}\\ 
	\end{block}
	\end{blockarray}
\label{eq:Pi}
	\end{equation}
to model the strategic competition between reactive and Bayesian strategies. To simplify our notations, we suppress the arguments of function $\pi$ (see Eq.~\ref{eq:Payoff_calculation}) wherever there is no ambiguity.

\section{Numerical Method}
In order to determine how a Bayesian strategy fares against an arbitrary reactive strategy, we allow the Bayesian player to play a repeated PD game against a large set of reactive strategies uniformly spread across the entire reactive strategy ($p$-$q$) space. The $p$-$q$ space is uniformly divided into $51\times51$ grid points. We numerically follow games played at every grid point. Three distinct kinds of interaction between the players at a grid point are possible, depending on the nature of the interacting partners: reactive-reactive, reactive-Bayesian, and Bayesian-Bayesian.

\begin{figure}[h]
	
	\centering
	\includegraphics[scale=0.73]{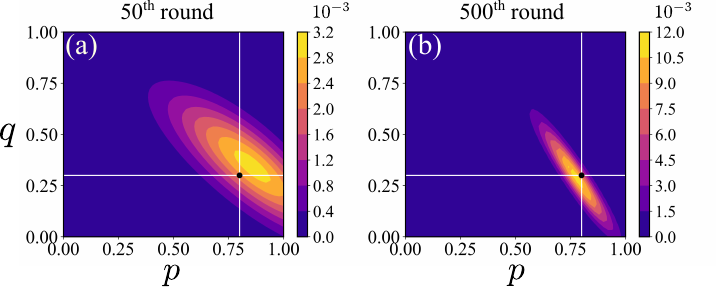}
	\caption{Convergence of belief:
		Subplots (a) and (b) respectively exhibit the posterior distribution of the Bayesian player's belief about the reactive player's strategy at the end of 50 rounds and 500 rounds, starting from a uniform prior. The black dot at the intersection of the horizontal line and the vertical line, represents the true reactive strategy, viz., $(0.8,0.3,0.5)$. The color bar indicates the probability, $P(p,q)$. The $p-q$ space has been divided into $51\times51$ grid points.}
	\label{fig:convergence}
\end{figure}

\begin{figure}[h]
	\centering
	\includegraphics[scale=0.7]{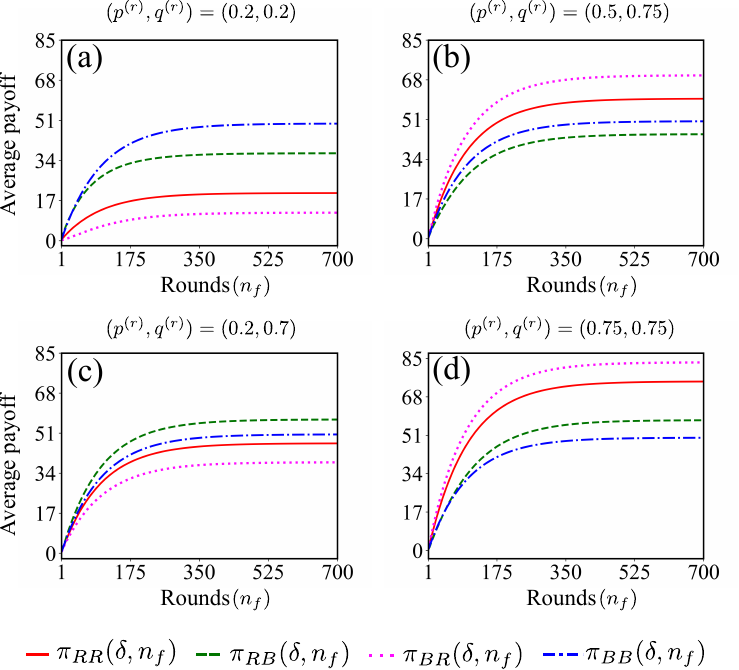}
	\caption{The variation of average payoff with total number of rounds: Subplots (a), (b), (c) and (d) correspond to the games where reactive strategy ($p^{(r)}$, $q^{(r)}$) is equal to $(0.2,0.2)$, $(0.5,0.75)$, $(0.2,0.7)$, and $(0.75, 0.75)$ respectively. Red solid, green dashed, pink dotted, and blue dot-dashed lines respectively denote $\pi_{RR}(\delta,n_f)$, $\pi_{RB}(\delta,n_f)$, $\pi_{BR}(\delta,n_f)$, and $\pi_{BB}(\delta,n_f)$. The cumulative payoffs are averaged over 10000 trials and computed for $r=2$. $\delta=0.99$ in all four cases.}
	\label{fig:payoff}
\end{figure}

When the focal player's strategy is reactive, i.e., $(p^{(r)},q^{(r)},c_1)$, we consider repeated games with reactive-reactive and reactive-Bayesian interactions to determine the payoff to the focal reactive player. When the focal player is Bayesian, games with both Bayesian-reactive and Bayesian-Bayesian interactions are considered to determine the payoff to the focal Bayesian player. We use an uniform prior distribution at $n=1$ represented by an $11\times11$ matrix where each matrix element is assigned equal probability. In other words, in our simulations, $p$ and $q$ are respectively sampled in steps of 0.1 from the ranges $0 \leq p \leq 1$ and $0 \leq q \leq 1$. The posterior distribution is then estimated at the end of each round using Eq.~(\ref{eq:Bayes}). Therefore, the posterior distribution also becomes a distribution over $11\times11$ grid points. Fig.~\ref{fig:convergence} illustrates the convergence of Bayesian player's belief, indicated by the peak of posterior distribution, towards the true reactive strategy. As expected, the convergence improves with the number of rounds. Each of the four types of interactions is repeated over 700 rounds to give the cumulative payoff for each pairwise interaction. The final payoffs $\pi_{RR}(\delta,n_f)$, $\pi_{RB}(\delta,n_f)$, $\pi_{BR}(\delta,n_f)$, and $\pi_{BB}(\delta,n_f)$; shown in Fig.~\ref{fig:payoff} are calculated by averaging over $10^4$ independent trials. Even though the time-evolution of the payoff is expected to be noisy, averaging over $10^4$ trials ensures smooth curves in Fig.~\ref{fig:payoff} that appear to be devoid of any fluctuations.


In passing, we remark that in line with the central limit theorem, the standard deviation about any of the average payoff elements is of the order $\sigma/\sqrt{10^{4}}$, where $\sigma$ is the standard deviation of the independent identical random payoffs. Since we have found that $\sigma\sim1$ implying $\sigma/\sqrt{10^{4}}\sim 10^{-2}$. Hence, we round off the average payoffs corresponding to each strategic interaction to the second place of decimal. Furthermore, we have fixed $c_1=0.5$ along the line of principle of insufficient reason~\cite{Jaynes2003}. We choose one  value of $\delta$ very close to unity (specifically, $\delta=0.99$) to somewhat negate the shadow of the future since, as we will see later, it facilitates certain analytical estimations. To see the effect of the discount factor $\delta$ on our simulation results, we compared results obtained for $\delta=0.75$ with those for $\delta=0.99$.

\section{Results}
Evidently, the reason behind the entire exercise of the aforementioned numerics is to calculate ${\sf\Pi}$ (see Eq.~\ref{eq:Pi}) for play at various points of the reactive strategy space. The central idea of this paper is to use these payoff matrices to ascertain the comparative efficiency of Bayesian strategy. To this end, one can envisage an unstructured population of randomly matched players. The success of the Bayesian strategy against any reactive strategy in a population is determined by the ability of the former to avoid being invaded by mutant reactive strategy. Therefore, we determine the conditions under which either the Bayesian strategy or the reactive strategy is an ESS and how those conditions are affected by the nature of the reactive strategy, the nature of the update rule (BTFT, NBTFT, or GBTFT) adopted by the Bayesian strategy, the benefit-to-cost ratio of cooperation and the discount factor ($\delta$). In such an investigation, however, we must distinguish between the cases of finite and infinite populations: While in the former case, a single mutant may invade the host monomorphic population; in the latter case, an infinitesimal fraction of mutants is required to invade a non-ESS host population strategy. Accordingly, the definition  of ESS varies in finite population~\cite{Nowak2004} and infinite population~\cite{Smith1973}. 

Therefore, in what follows, we succinctly  present our results using, what we term as, ESS phase diagram. An ESS phase diagram is a pictorial representation showing which strategy is an ESS in which region of the reactive strategy space. The regions with Bayesian strategy as an exclusive ESS, reactive strategy as an exclusive ESS, and both the strategies as ESS are marked with different colours. Whereas, the region where mixed ESS exists i.e., certain non-zero fraction of the population plays Bayesian and the rest non-zero fraction adopts reactive strategy, we mark it using a colour gradient such that its intensity (redness) denotes the frequency of reactive strategy. Obviously, the mixed ESS should be absent in the finite population scenario.
 \subsection{Infinite population}
 Before we start discussing the ESS phase diagrams generated from our simulation, let us recall the condition of ESS for a given payoff matrix $\sf{\Pi}$~\cite{Smith1982}: The reactive strategy is ESS in an infinite population if (a) $\pi_{RR}>\pi_{BR}$ or (b) $\pi_{RR}=\pi_{BR}$ and $\pi_{RB}>\pi_{BB}$; similarly, The Bayesian strategy is ESS if (a) $\pi_{BB}>\pi_{RB}$ or (b) $\pi_{BB}=\pi_{RB}$ and $\pi_{BR}>\pi_{RR}$. Finally, a mixed ESS is implied by the condition: $\pi_{RR}<\pi_{BR}$ and $\pi_{BB}<\pi_{RB}$.
  
 \begin{figure}[h!]
 	\centering
 	\includegraphics[scale=1.1]{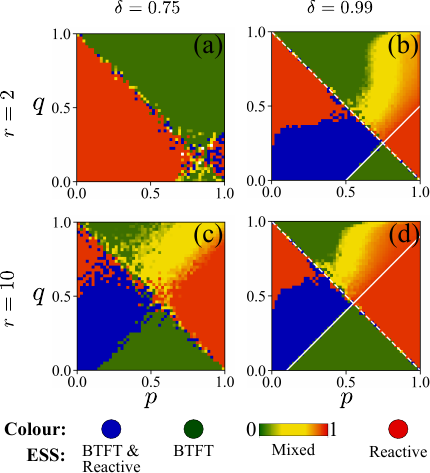}
 	\caption{ESS phase diagram for BTFT vs. reactive strategies in infinite population: Four subplots (a), (b), (c), and (d) respectively correspond to $r=2$ and $\delta=0.75$,  $r=2$ and $\delta=0.99$, $r=10$ and $\delta=0.75$, and $r=10$ and $\delta=0.99$. The color gradient, ranging from zero to unity, marks the frequency of reactive strategy in the mixed ESS. Blue, green, and red colors, respectively, represent three cases: BTFT and reactive strategy are both ESS's, BTFT is the only ESS, and reactive is the only ESS. Few scattered white dots correspond to absence of any ESS. The dashed white line and the solid white line---the analytically estimated boundaries---respectively satisfy equations: $q=1-p$ and  $q=p-1/r$.} 
 	\label{fig:BTFT}
 \end{figure}
	\subsubsection{BTFT}
In this subsection, we consider the competition between the reactive and BTFT strategy and analyse the
ESS phase diagram for a given discount factor and benefit-to-cost ratio. It is clear that diagram depends on the nature of the reactive strategy competing with the Bayesian strategy. For low values of the benefit-to-cost ratio ($r$) and the discount factor ($\delta$) (see Fig.~\ref{fig:BTFT}(a)), a host population of reactive strategy players can prevent invasion by an infinitesimal fraction of a mutant Bayesian strategy as long as $p+q<1$ and $p\lesssim 0.75$. Similarly, the Bayesian strategy is an ESS, and can therefore prevent invasion by any reactive strategy, as long as $p+q>1$ and $q \gtrsim 0.25$. In this region, a BTFT mutant can invade a reactive strategy.

Increasing either the benefit-to-cost ratio ($r$) or the discount factor ($\delta$), or both, have a similar effects as can be seen by comparing Fig.~\ref{fig:BTFT}b--\ref{fig:BTFT}d with Fig.~\ref{fig:BTFT}a. Moreover, for a certain range of $(p,q)$ values lying in the region $p+q<1$, both reactive and Bayesian strategies are an ESS but the size and location of this region varies as both $r$ and $\delta$ changes. Another region shown in yellow corresponding to $p+q>1$ is characterized by the stable coexistence between the reactive and BTFT strategy when neither of the strategies is an exclusive ESS; rather a mixed ESS exists.

During the process of inferring the reactive opponent's true strategy, the Bayesian player samples many different $(p,q)$ values as it acquires evidence based on the opponent's actions. While the region of reactive space from where $(p,q)$ values are sampled becomes eventually restricted as evidence accumulates over increasing number of rounds; initially, when evidence is sparse, the region can be large. The exact stochastic trajectories of the Bayesian player in the reactive space is analytically intractable. However, given the sharp phase boundaries in ESS phase diagrams, an explanation about why they appear as they do is worth uncovering. Moreover, there are a few additional intriguing features of the ESS phase diagram that are worth understanding, e.g., ALLD may not invade BTFT (see Fig.~\ref{fig:BTFT}b--\ref{fig:BTFT}d) and ALLC is not completely eliminated by a mutant BTFT but can coexist with the latter.

To this end, we present a very interesting {\it ansatz}, in the limit $\delta\to1$: It is helpful to think of the BTFT as an effective reactive strategy with $p=q=0.5$, once again invoking the principle of insufficient reason. Even though this ansatz cannot explain all aspects of the complex, evolutionary dynamics between the Bayesian and reactive strategies, it is successful in explaining aforementioned specific features of the dynamics. In this context, a caveat is worth pointing out. It should be noted that if $n_f\to\infty$, the average payoff can be written as $(1-\delta)\sum_{n=1}^{n_f}u_n\delta^{n-1}$ for the discount factor $\delta\in(0,1)$. However, if $\delta=1$, then the average payoff is  $\lim_{n_f\rightarrow\infty}({\sum_{n=1}^{n_f}u_n}/{n_f})$ (when the limit exists). In what follows we work with the premise that the two definitions of average payoff corresponding to $\delta\rightarrow1$ and $\delta=1$ are equivalent for sufficiently large enough $n_f$ and for $\delta$ sufficiently close to 1. Consequently, to find ${\sf \Pi}$, we merely need to recall the standard result~\cite{Nowak1990} that when two players---each with two reactive strategies $S_1\equiv(p_1,q_1)$ and $S_2\equiv(p_2,q_2)$ play against each other, the payoff elements are given by
\begin{equation}
	\pi_{S_iS_j}=(r-1)c^\infty_1c^\infty_2-c^\infty_1(1-c^\infty_2)+r(1-c^\infty_1)c^\infty_2,\label{eq:effective}
\end{equation}
 $\forall i,j\in\{1,2\}$. Here the superscript $\infty$ represents the limiting stationary cooperation probability given by $c_i^{\infty}=q_i/(1-p_i+q_i)$ for $i=1,2$.
 
Now we are equipped to explain the features pointed out earlier. First, let us focus on the phase boundaries. We have at any grid point of the reactive space, two strategies: $R=(p,q)$ and $B=(1/2,1/2)$. The reactive strategy $R$ is an ESS if $\pi_{RR}> \pi_{BR}$ which, owing to Eq.~(\ref{eq:effective}), leads to inequality $r(p-q)(p+q-1)>(p+q-1)$, i.e.,
\begin{subequations}
	\begin{eqnarray}
&&q<p-\frac{1}{r}~{\rm if}~p+q>1,\\
&&q>p-\frac{1}{r}~{\rm if}~p+q<1.
	\end{eqnarray}
\label{eq:diag}
\end{subequations} 
This estimation is very promising as evident from Fig.~\ref{fig:BTFT}(b) and Fig.~\ref{fig:BTFT}(d): The regions (red and blue) with reactive strategy as ESS satisfy inequalities~(\ref{eq:diag}). The lines $p+q=1$ (dashed white) and $q=p-1/r$ (solid white) are almost precise estimates of the phase boundaries.

The evolutionary stability of the ALLC ($p = 1, q = 1$) and ALLD ($p = 0, q = 0$) strategies against the Bayesian player depends on both $r$ and $\delta$, as is evident from Fig.~\ref{fig:BTFT}. We first consider the game between ALLC and BTFT. In the $\delta \rightarrow 1$ limit, coexistence is observed between ALLC and Bayesian strategy since neither is an ESS (see Fig.~\ref{fig:BTFT}b,d). When ALLC cooperates in the first round, the likelihood estimated by the Bayesian player is $P(C|p,q)=(p+q)/2$ which is maximum at $(p,q)=(1,1)$. Since we start with a uniform prior, the posterior probability $P(p,q|C)$ used by the BTFT player to infer the strategy of ALLC opponent also peaks at $(p,q)=(1,1)$. ALLC cooperates in all subsequent rounds rendering the belief of BTFT player fixed at $(p,q)=(1,1)$. Hence, BTFT converges to the ALLC right after the first round, i.e., the BTFT behaves as an effective reactive strategy with $p=q=1$; her cumulative payoff is, thus, $\pi_{BR}=r-1$. Clearly, $\pi_{RB}=\pi_{BR}=r-1$. However, recall that by construction ALLC can also defect in the first round with a probability of $0.5$. Under such circumstances
the  Bayesian player's belief cannot converge quickly to $(p,q)=(1,1)$; her belief tracks a stochastic trajectory in $p$-$q$ space. Invoking the ansatz that BTFT effectively act likes a reactive strategy with $(p,q)=(1/2,1/2)$, we get $\pi_{BR}=(2r-1)/2$ and $\pi_{RB}=(r-2)/2$ using Eq.~(\ref{eq:effective}). Ergo, the effective $\pi_{BR}$ should have contributions from the payoffs obtained by BTFT while acting as reactive strategy $(p,q)=(1/2,1/2)$ and reactive strategy $(p,q)=(1,1)$. Since $c_1=0.5$, we consider that these contributions come with \textit{equal} weights; and hence, the final effective $\pi_{BR}=[0.5(2r-1)/2]+0.5(r-1)=r-{3}/{4}$. Similarly,  $\pi_{RB}=[0.5(r-2)/2]+0.5(r-1)=({3r}/{4})-1$ respectively.

In order to show the coexistence of ALLC and BTFT, we must show that neither ALLC nor BTFT is an ESS i.e. 
$\pi_{RR}<\pi_{BR}$ and $\pi_{BB}<\pi_{RB}$. While we just calculated $\pi_{BR}$ and $\pi_{RB}$, $\pi_{RR}=(r-1)$ but $\pi_{BB}$ remains to be estimated. When two BTFT players play against each other, their payoffs depend on whether the action profile in the very first round is $(C,C)$, $(D,D)$, or $(C,D)$ (equivalently, $(D,C)$) which corresponds to the Bayesian players effectively playing ALLC, ALLD and reactive strategy with $(p,q)=(0.5,0.5)$ respectively. Since $c_1=0.5$, the three forms of the BTFT should be respectively associated with weight factors 0.25, 0.25 and 0.5 for the calculation of the payoff. The weighted payoff, thus, is given by $\pi_{BB}=0.25(r-1)+0.25(0)+0.5(r-1)/2=(r-1)/2$. Evidently, $\pi_{RR}<\pi_{BR}$ always holds good, whereas $\pi_{BB}<\pi_{RB}$ implies $r>2$. Of course, this is not a strict bound given the non-rigorous assumptions made about the BTFT's dynamics and the mean-field nature of the arguments. Nevertheless, it is remarkable, how we arrive at a condition on the benefit-to-cost ratio (keeping $\delta\to1$) for which the coexistence of ALLC and BTFT is a distinct possibility.

Finally, the case of ALLD vs. BTFT may be treated in a similar manner to show that ALLD may not invade BTFT. While playing against ALLD, the BTFT acts like ALLD after the first round if ALLD defects in the first round (since the likelihood function $P(D|p,q)=(2-p-q)/2$ is maximum at $p=q=0$) and she plays like a reactive strategy with $(p=0.5,q=0.5)$ if ALLD cooperates in the first round. The weight factors associated with these two roles of BTFT are each equal to 0.5. Hence the weighted payoffs are $\pi_{BR}=-1/4$ and $\pi_{RB}={r}/{4}$, and $\pi_{RR}=0$. $\pi_{BB}=(r-1)/2$ was already estimated in the preceding paragraph. ALLD is an ESS if $\pi_{RR} > \pi_{BR}$ which is trivially satisfied and the Bayesian strategy is an ESS if $\pi_{BB} > \pi_{RB}$ which is satisfied for $r>2$. This accounts for the observation (see Fig.~\ref{fig:BTFT}b and Fig.~\ref{fig:BTFT}d) that both ALLD and Bayesian strategies are ESS's.
	\subsubsection{NBTFT}
\begin{figure}
	\centering
	\includegraphics[scale=0.57]{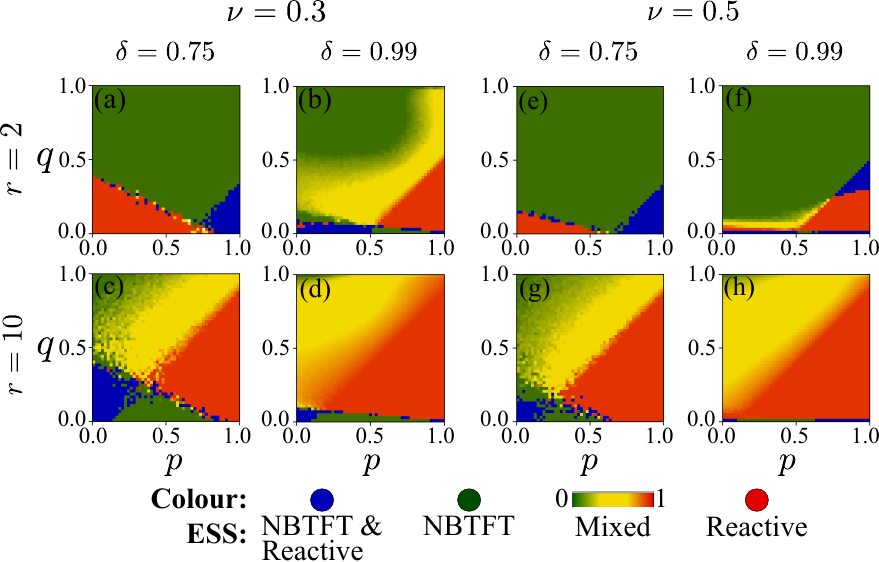}
	\caption{ESS phase diagram for  NBTFT vs. reactive strategies in infinite population: Two sets of subplots, viz., $\{\rm{a,b,c,d}\}$ and $\{\rm{e,f,g,h}\}$ respectively correspond to two non-reciprocity parameter values $\nu=0.3$ and $\nu=0.5$. For each set, four subplots are arranged in a $2\times2$ grid corresponding to two different discount factors $\delta\in\{0.75,0.99\}$ and two benefit-to-cost ratio parameter values $r\in\{2,10\}$. The color gradient, ranging from zero to unity, marks the frequency of reactive strategy in the mixed ESS. Blue, green, and red colors, respectively, represent three cases: BTFT and reactive strategy are both ESS's, BTFT is the only ESS, and reactive is the only ESS. Few scattered white dots correspond to absence of any ESS.}
	\label{fig:NBTFT}
\end{figure}

	When the Bayesian player chooses not to reciprocate fully, she  modifies her strategy to an NBTFT strategy that results in her cooperating with probabilities ($p^{\prime},q^{\prime}$) that are {\it lower} than her belief ($p^{\rm max},q^{\rm max}$) about opponent's $(p,q)$ value. This way she can potentially exploit the reactive opponent more frequently and thereby acquire a larger cumulative payoff. Hence such a strategy is able to resist invasion by a reactive counterpart over a larger region of reactive strategy space as can be seen by comparing the cases of $r=2$ in Fig.~\ref{fig:NBTFT} with that in Fig.~\ref{fig:BTFT}. Here the region of dominance of NBTFT strategy increases with increasing $\nu$ which quantifies the extent of non-reciprocity of the NBTFT strategy compared to the BTFT strategy (Fig.~\ref{fig:NBTFT}a,b vs. Fig.~\ref{fig:NBTFT}e,f). This advantage is most pronounced for higher $(p,q)$ values and less effective for reactive opponents with low $(p,q)$ values. 
	
	However, the advantages of NBTFT decreases as the benefit-to-cost ratio of cooperation increases (compare upper panels with lower panels of Fig.~\ref{fig:NBTFT}). Specifically, we observe that for $r=10$, NBTFT is ESS only in regions characterized by smaller values of both $(p,q)$ and small $p$, large $q$. This is because the much larger benefit for mutual cooperation (compared to mutual defection) that accrues over time outweighs the occasional advantage of selfish behaviour exhibited by the NBTFT player. Reactive strategies with large $q$ are more prone to exploitation by NBTFT since they have a higher likelihood of cooperating even when the NBTFT opponent defects.

With increasing $\delta$, the contributions to payoff from later rounds carry almost as much weight as the contributions from earlier rounds.  It is also important to note that the NBTFT player keeps updating her belief and as her belief converges towards the true belief about her reactive opponent with increasing number of rounds, her ability to exploit the cooperative nature of her reactive opponent is limited to those reactive strategies with higher $q$ values (compare Fig.~\ref{fig:NBTFT}a and Fig.~\ref{fig:NBTFT}b, Fig.~\ref{fig:NBTFT}c and Fig.~\ref{fig:NBTFT}d). The dominance of NBTFT strategies is lost when both $r$ and $\delta$ increase (see Fig.~\ref{fig:NBTFT}d and Fig.~\ref{fig:NBTFT}h) indicating that increased benefits from mutual cooperation over mutual defection as well as enhanced contribution to total payoff from later rounds increasingly favour reactive strategies with $q<p$, leading to their dominance for $q<p$ (see Figs.~\ref{fig:NBTFT}d and \ref{fig:NBTFT}h) and coexistence of NBTFT and reactive strategies is seen only for $q>p$ (see Figs.~\ref{fig:NBTFT}d and \ref{fig:NBTFT}h). NBTFT is found to dominate only in a small sliver of region around $q=0$ and $q=1$.

	\subsubsection{GBTFT}
		\begin{figure}
		\centering
		\includegraphics[scale=0.57]{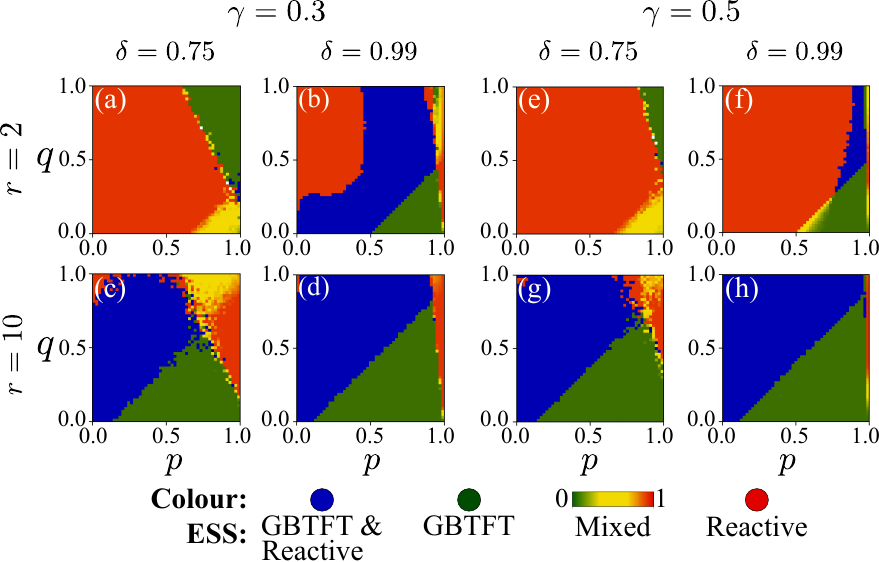}
		\caption {ESS phase diagram for  GBTFT vs. reactive strategies in infinite population: Two sets of subplots, viz., $\{\rm{a,b,c,d}\}$ and $\{\rm{e,f,g,h}\}$ respectively correspond to two non-reciprocity parameter values $\gamma=0.3$ and $\gamma=0.5$. For each set, four subplots are arranged in a $2\times2$ grid corresponding to two different discount factors $\delta\in\{0.75,0.99\}$ and two benefit-to-cost ratio parameter values $r\in\{2,10\}$. The color gradient, ranging from zero to unity, marks the frequency of reactive strategy in the mixed ESS. Blue, green, and red colors, respectively, represent three cases: BTFT and reactive strategy are both ESS's, BTFT is the only ESS, and reactive is the only ESS. Few scattered white dots correspond to absence of any ESS.}
		\label{fig:GBTFT}
	\end{figure} 

If the strategic response is more generous than BTFT (i.e. $p'_{n+1}>p^{\rm max}_{n}$ and $q'_{n+1}>q^{\rm max}_{n}$), the corresponding GBTFT player is easily exploited by the reactive opponent for most values of $(p,q)$ when both $r$ and $\delta$ are small (compare Fig~\ref{fig:BTFT}a and Fig.~\ref{fig:GBTFT}a). GBTFT can resist invasion by a reactive strategy only if the reactive opponent's are highly cooperative, i.e. both $p$ and $q$ are sufficiently high (see Fig.~\ref{fig:GBTFT}a). As expected, the region of $(p,q)$ space where GBTFT is an ESS shrinks even further as $\gamma$, which quantifies the extent of generosity shown by the GBTFT player, increases (compare Fig.~\ref{fig:GBTFT}a and Fig.~\ref{fig:GBTFT}e).

	As the benefit-to-cost ratio $r$ of cooperation increases, more cooperative strategies gain an advantage from the larger payoff received for mutual cooperation which offsets the cost of being exploited by the opponent's selfish behaviour. For this reason, reactive strategies with high $(p,q)$ outperform their Bayesian counterpart and are therefore stable against invasion by GBTFT (see Fig.~\ref{fig:GBTFT}c). On the other hand, GBTFT dominates over the reactive counterparts when the probability of reciprocal cooperation, $p$, for reactive strategy is above a certain threshold value of $q$: $p=q+\epsilon$, $\epsilon \approx 0.15$ ($r=10$ and $\delta=0.75$). For such types of reactive opponent's, GBTFT can reap the large benefit of mutual cooperation while occasionally exploiting the cooperative nature of the reactive opponent. A region of reactive strategy space where both strategies are ESS's also emerges (blue region in Fig.~\ref{fig:GBTFT}c).

With an increase in the discount factor ($\delta$), the region of $(p,q)$ space where GBTFT dominates changes to one characterized by high $p$ and low $q$. This region increases with increased benefit-to-cost ratio (compare Fig.~\ref{fig:GBTFT}b with Fig.~\ref{fig:GBTFT}d) and the generosity parameter $\gamma$ (compare with Fig.~\ref{fig:GBTFT}d and Fig.~\ref{fig:GBTFT}h) since the enhanced advantage of mutual cooperation carry more weight over larger time scales (due to larger $\delta$).
 
\subsection{Finite population}
\begin{figure*}
	\centering
	\includegraphics[scale=0.82]{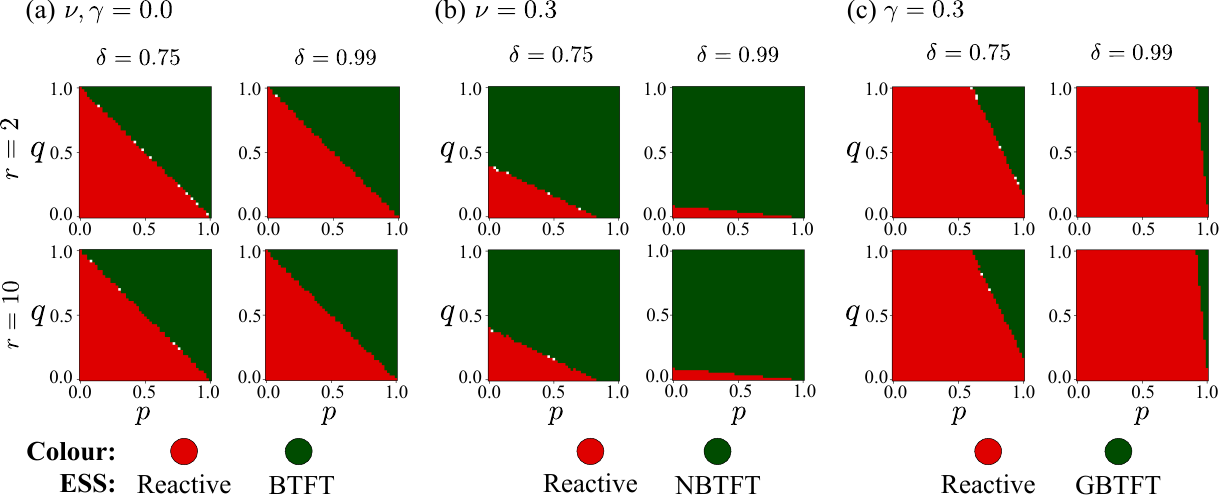}
	\caption{ESS phase diagram for a repeated 2-player game between the Bayesian strategies and the reactive strategies for a finite population of size $N=2$: Subplots (a), (b), and (c) represent three Bayesian strategies; BTFT, NBTFT with non-reciprocity parameter $\nu=0.3$, and GBTFT with generosity parameter $\gamma=0.3$, respectively. In each subplot, the red color denotes that the reactive strategy is an ESS, green color denotes that the Bayesian strategy is ESS. Few scattered white dots correspond to absence of any ESS for those ($p,q$) values.} 
	\label{fig:N2}
\end{figure*}
\begin{figure}[h]
	\centering
	\includegraphics[scale=0.42]{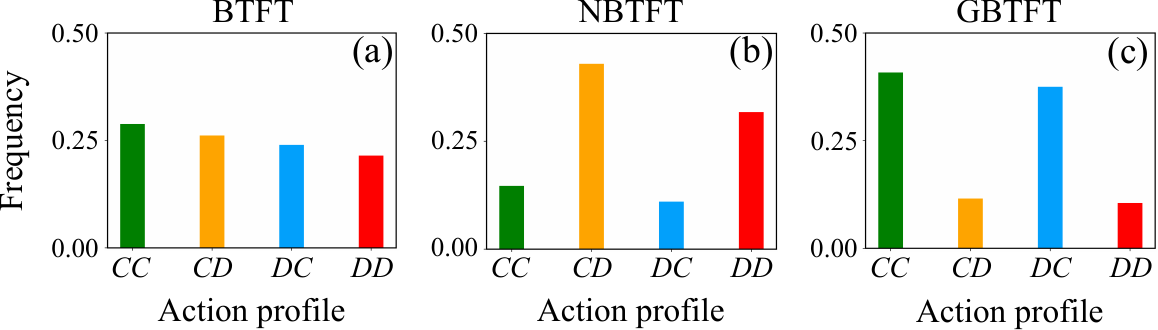}
	\caption{Histogram of different interactions generated in a repeated two-player game with reactive and Bayesian strategies: Subplots (a), (b), and (c), respectively, correspond to the interaction of the three Bayesian strategies; BTFT, NBTFT (with $\nu=0.5$) and GBTFT (with $\gamma=0.5$) with a reactive strategy defined by $(0.5,0.6,0.5)$.  The game is played over $700$ rounds and the frequency of each interaction pair is obtained by averaging over $10^4$ distinct trials. The first and the second actions correspond to those of the reactive and the Bayesian player respectively.}
	\label{fig:INDIA}
\end{figure}
\begin{figure}
	\centering
	\includegraphics[scale=0.6]{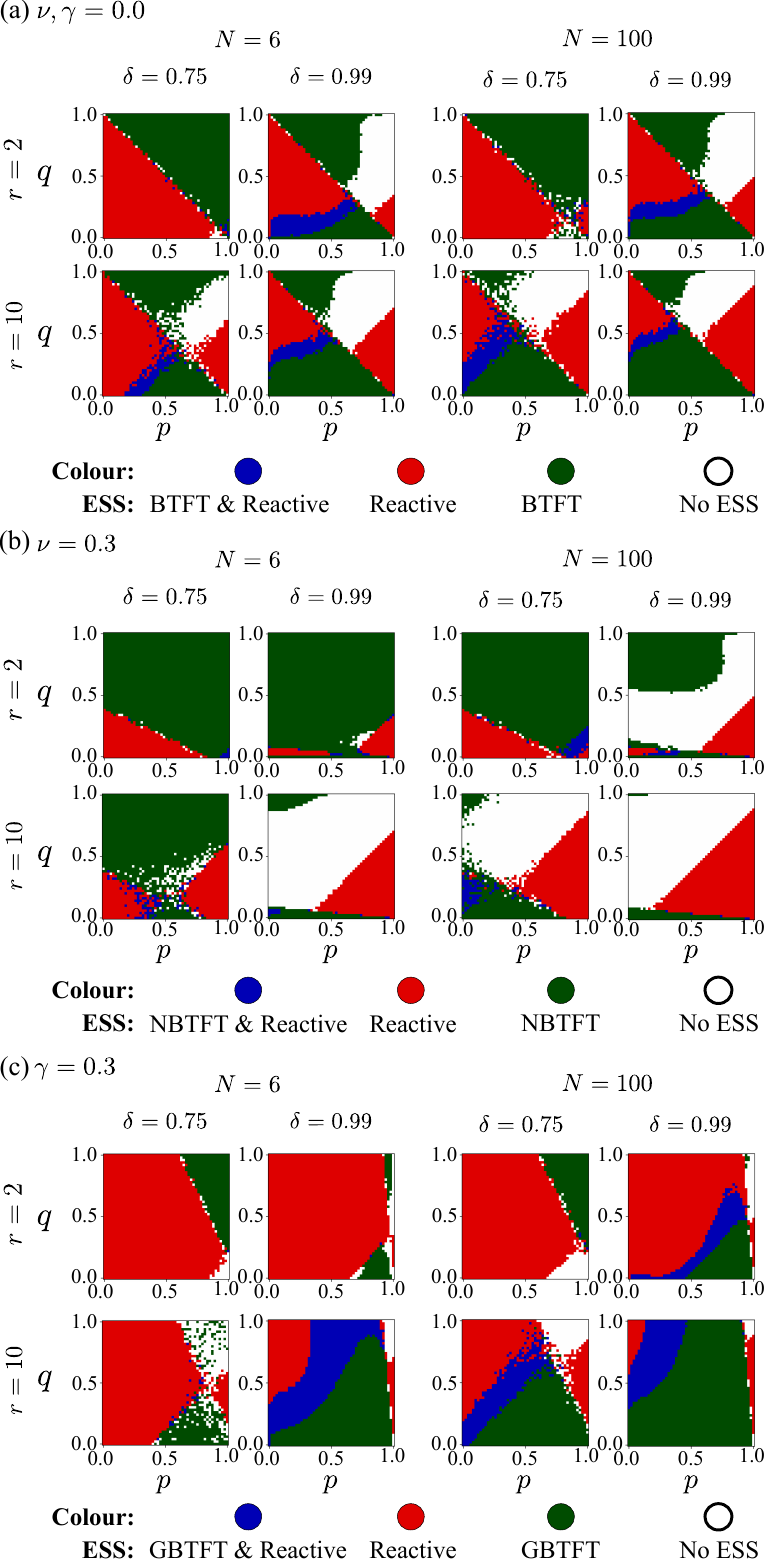}
	\caption{ESS phase diagram for  Bayesian strategies vs. reactive strategies in finite populations of sizes $N=6$ and $N=100$: Subplots (a), (b), and (c) represent three Bayesian strategies; BTFT, NBTFT with non-reciprocity parameter $\nu=0.3$, and GBTFT with generosity parameter $\gamma=0.3$, respectively. In each subplot, while red color denotes that reactive strategy is ESS, green color denotes that Bayesian strategy is ESS. The white color corresponds to absence of any ESS.} 
	\label{fig:N100}
\end{figure}
In finite populations, evolutionary stability of a Bayesian strategy is dependent on the population size $N$. Consequently, the ESS condition needs to be appropriately modified~\cite{Nowak2004} to ensure that a single reactive mutant has a lower fitness than the Bayesian strategy {\it and} selection opposes the fixation of reactive mutant, i.e., the fixation probability $\rho_R$ of reactive strategy is less than $1/N$. The former condition leads to $(N-1)\pi_{RB}<\pi_{BR}+(N-2)\pi_{BB}$ while the latter leads to $(N-2)\pi_{RR}+(2N-1)\pi_{RB}<(N+1)\pi_{BR}+(2N-4)\pi_{BB}$ under the assumption of weak selection. Similarly, the evolutionary stability of a reactive strategy implies that the conditions $(i)(N-1)\pi_{BR}<\pi_{RB}+(N-2)\pi_{RR}$ and $(ii)(N-2)\pi_{BB}+(2N-1)\pi_{BR}<(N+1)\pi_{RB}+(2N-4)\pi_{RR}$ are simultaneously satisfied. In finite populations, there are only two absorbing states corresponding to a population consisting only of either the Bayesian strategy or the reactive strategy. Hence a mixed phase where both strategies coexist is not possible.

For $N=2$, the game is played between a single Bayesian and a single reactive player, and the aforementioned conditions of ESS simply boils down to  $\pi_{RB}>\pi_{BR}$ for the reactive strategy to be an ESS and $\pi_{BR}>\pi_{RB}$ for the Bayesian strategy to be an ESS. In other words, the condition for evolutionary stability depends on which of the two strategies has a larger average payoff when playing against the other. From Fig.~\ref{fig:N2}, it is clear that reactive strategies dominate over their Bayesian (BTFT) counterparts as long as $p+q<1$. This can be rationalized through the ansatz that the Bayesian strategy may be thought as an effective reactive strategy with $p=q=0.5$. Thus, as before, calculations yield that $\pi_{RB}=(r-p-q)/2$ and $\pi_{BR}=[r(p+q)-1]/2$ where $(p,q)$ denote the opponent's reactive strategy. Thus, the condition for the reactive strategy to beat the Bayesian one gets recast as $(r-p-q)/2>[r(p+q)-1]/2$ which implies the condition $p+q<1$. Interestingly, the condition is independent of the benefit-to-cost ratio $r$ as observed from Fig.~\ref{fig:N2}. 

When the Bayesian strategy chooses to be more selfish (NBTFT) than is dictated by her perceived belief about their reactive opponent's strategy, it dominates over the reactive strategy over a much larger region of $(p,q)$ space (Fig.~\ref{fig:N2}b). As $\delta$ increases, the size of this region increases with NBTFT dominating all but the most selfish strategies (see Fig.~\ref{fig:N2}b). The NBTFT player, by virtue of her ability to explore the strategy space in the process of inferring her opponent's strategy, is more effective in exploiting her reactive opponent (see Fig.~\ref{fig:N2}b), leading to a larger average payoff for herself. The situation is reversed when the Bayesian strategy is GBTFT  (see Fig.~\ref{fig:N2}c), i.e, more generous. The reactive strategy dominates over a larger region of strategy space as $\delta$ increases, indicating that it is better able to exploit the generosity of the BTFT strategy (see Fig.~\ref{fig:N2}c) to increase her average payoff. 

This observations can be explained by noting that in contrast to the BTFT case (comparing Fig.~\ref{fig:INDIA}a and ~\ref{fig:INDIA}b); the fewer average number of C-C interactions and larger average number of D-D interactions between the NBTFT and the reactive player ensures that higher benefits of mutual cooperation does not accrue as much. Similarly, comparing Fig.~\ref{fig:INDIA}a and ~\ref{fig:INDIA}c, we note that even though the number of C-C interactions is larger in the latter case, the significantly larger (on average) number of DC interactions (indicative of the reactive player more frequently exploiting the GBTFT player) neutralizes the increased benefits of mutual cooperation. 

As the population size increases, the ESS phase diagram in $(p,q)$ space approaches the infinite population limit as can be seen by comparing the $N=100$ case in Figs.~\ref{fig:N100}a--c with the corresponding panels in Figs.~\ref{fig:BTFT}--\ref{fig:GBTFT}. It should be noted the mixed ESS (coexisting reactive and Bayesian players) does not exist in the finite population scenario because there are two absorbing states which can be an ESS. Hence, wherever there is a mixed ESS in the infinite population case, a white region (No ESS) appears in the corresponding ESS phase diagram for finite populations. 
	
\section{Discussion}
Can a strategy that attempts to learn the fixed reactive strategy of the opponent prevent being out-competed by extremely selfish strategies like ($p \sim 0,q \sim 0$) in a repeated PD game? The answer depends critically on the relative benefit of cooperation ($r$), the discount factor ($\delta$) and on the nature of the strategy (BTFT, GBTFT, or NBTFT) employed by the Bayesian player to update her actions. For low $r$ and $\delta$, predominantly selfish strategies dominate over Bayesian learning strategies (see Figs.~\ref{fig:BTFT}a,~\ref{fig:NBTFT}a,and \ref{fig:GBTFT}a). But the situation changes with increase in the discount factor (see Figs.~\ref{fig:BTFT}b,~\ref{fig:NBTFT}b,and \ref{fig:GBTFT}b). As $r$ increases, the Bayesian learning strategies are always effective at resisting invasion by selfish reactive strategies even for low discount factors (see Figs.~\ref{fig:BTFT}c,~\ref{fig:NBTFT}c,and \ref{fig:GBTFT}c). 

Even though the Bayesian player may end up being more cooperative than extremely selfish strategies during the exploration phase of the game when she is trying to learn the strategy of her opponent, she avoids exploitation in the long run by gradually becoming more selfish through effective learning of her opponent's strategy. In general, the success of a Bayesian player depends on the extent to which she can leverage the higher benefits of mutual cooperation against a cooperative opponent while avoiding being exploited by a more selfish opponent.

Reactive strategies form just a subset of the larger class of Markovian memory-one strategies. Our results can be easily extended to see how Bayesian strategies fare against more cognitively sophisticated Markovian strategies~\cite{Minjae2021}. Bayesian inference in evolutionary games provides a powerful learning framework applicable to other social dilemmas that can be modeled through the public goods game. In such situations each individual will take into account the actions of other members of her community to update her belief about cooperation levels of the group and tune her actions accordingly. It would be interesting to see how Bayesian learning compares with other strategy update mechanisms like pairwise comparison and reinforcement learning in such scenarios.

In order to better understand the key underlying causes behind altruistic behaviour in the natural world, it is important to take into account realistic ways in which animals learn and take decisions. Decision making is often modulated by learning as well as cognitive constraints in factoring and processing a diverse range of stimuli from the environment. Accounting for those constraints will enable us to build more realistic models for understanding altruistic behaviour in social groups. The Bayesian framework developed in this work is the first step in incorporating sophisticated statistical learning mechanisms like Bayesian learning in altruistic decision-making. We hope to eventually address scenarios in which deviations from Bayesian inference, perhaps induced by cognitive constraints, can also affect patterns of altruistic behaviour in social groups. Such investigations will hopefully make it possible to design and implement protocols that encourage altruistic behaviour leading to greater benefits for society at large.

\acknowledgements
CSIR (India) is acknowledged for awarding a senior research fellowship to AP. SC and SS acknowledge the support from SERB (Govt. of India) through projects no. MTR/2021/000119 and MTR/2020/000446 respectively.  

\bibliography{patra_etal_references}    
\end{document}